\documentclass[prd,aps,preprint,epsfig,floats]{revtex4}
\usepackage{graphics}
\usepackage{epsfig}

\usepackage{graphicx}

 \preprint{UFIFT-HEP-06-20}\preprint{OSU-HEP-06-14}

\begin{document}
\title{
Unified TeV Scale Picture of Baryogenesis and Dark Matter }

 \author{K. S. Babu $^{\it \bf a}$~\footnote[3]
{Email:kaladi.babu@okstate.edu}}

 \author{R. N. Mohapatra $^{\it \bf b}$~\footnote[2]
{Email:rmohapat@physics.umd.edu}}

 \author{Salah Nasri $^{\it \bf c}$~\footnote[4]
{Email:snasri@phys.ufl.edu}}

 \affiliation{$^ {\bf \it a}$Department of Physics, Oklahoma State University,
Stillwater, OK 74078, USA}

\affiliation{$^ {\bf \it b}$Department of Physics, University of
Maryland,College Park, MD 20742,USA}

\affiliation{$^{\bf \it c}$ Department of Physics, University of
Florida, Gainesville, Florida 32611, USA}


\begin{abstract}
We present a simple extension of MSSM which provides a unified
picture of cosmological baryon asymmetry and dark matter. Our model
introduces a gauge singlet field $N$ and a color triplet field $X$
which couple to the right--handed quark fields. The out--of
equilibrium decay of the Majorana fermion $N$ mediated by the
exchange of the scalar field $X$ generates adequate baryon asymmetry
for $M_N \sim 100$ GeV and $M_X \sim$ TeV. The scalar partner of $N$
(denoted $\tilde{N}_1$) is naturally the lightest SUSY particle as
it has no gauge interactions and plays the role of dark matter.
$\tilde{N}_1$ annihilates into quarks efficiently in the early
universe via the exchange of the fermionic $\tilde{X}$ field. The
model is experimentally testable in (i) neutron--antineutron
oscillations with a transition time estimated to be around $10^{10}$
sec, (ii) discovery of colored particles $X$ at LHC with mass of
order TeV, and (iii) direct dark matter detection with a predicted
cross section in the observable range.
\end{abstract}

\maketitle

\section{ Introduction}. The origin of matter--anti-matter
asymmetry of the Universe and that of dark matter are two of the
major cosmological puzzles that rely heavily on particle physics
beyond the standard model for their resolution. It is a common
practice to address these two puzzles separately by invoking
unrelated new physics. For instance, a widely held belief is that
either the lightest supersymmetric particle (LSP) or the near
massless invisible axion constitutes the dark matter, while
baryogenesis occurs through an unrelated mechanism involving either
the decay of a heavy right--handed neutrino (leptogenesis), or new
weak scale physics which makes use of the electroweak sphalerons. A
closer examination of the minimal versions of SUSY would suggest
that to generate the required amount of dark matter density one
needs some tuning of parameters. The LSP should either have the
right amount of Higgsino component, or another particle, usually the
right--handed stau, should be nearly degenerate with the Bino LSP to
facilitate dark matter co-annihilation. Similarly, the leptogenesis
mechanisms requires the heavy right--handed neutrino to have its
mass in the right range to generate the adequate amount of matter.
Despite these possible problems, these ideas are attractive since
they arise in connection with physics scenarios which are strongly
motivated by other puzzles of the standard model, e.g., resolving
the gauge hierarchy problem (in the case of LSP dark matter), or
generating small neutrino masses (in the case of leptogenesis). In
the absence of any experimental confirmation of these ideas, it is
quite appropriate to entertain alternate explanations which could be
motivated on other grounds. Our motivation here is to seek a unified
picture of both these cosmological puzzles within the context of
weak scale supersymmetry without fine--tuning of parameters. We
propose a class of models where a very minimal extension of the MSSM
resolves these puzzles in a natural manner with testable
consequences for the near future.

Our extension of MSSM involves the addition of two new particles: a
SM singlet superparticle denoted by  $N$ with mass in the 100 GeV
range and an iso-singlet color triplet particle $X$ with mass in the
TeV range.  These particles, consistent with the usual $R$--parity
assignment, couple only to the right--handed quark fields. We
discuss two models, one in which the electric charge of $X$ is $2/3$
and another where it is $-1/3$. We show that in these models, baryon
asymmetry arises by the mechanism of post--sphaleron baryogenesis
suggested by us in a recent paper \cite{bmn} involving the decay of
the Majorana fermion $N$. The scalar component of $N$ (denoted as
$\tilde{N}_1$) has all the right properties to be the cold dark
matter of the universe without any fine--tuning of parameters.  The
purpose of the heavier $X$ particle is to facilitate baryon number
violation in the interaction of $N$, and also to help $\tilde{N}_1$
annihilate into quarks. A very interesting prediction of these
models is the existence of the phenomenon of neutron-anti-neutron
oscillation with a transition time in the accessible range of around
$10^{10}$ sec. The TeV scale scalar $X$  and its fermionic
superpartner $\tilde{X}$ are detectable at LHC. Furthermore, the
model predicts observable direct detection cross section for the
dark matter.

\section{Outline of the model}. As already noted, we add two new
superfields to the MSSM -- a standard model singlet $N$ and a pair
of color triplet with weak hypercharge $=\pm 4/3$ denoted as
$X,\overline{X}$. The $R$--parity of the fermionic component of $N$
is even, while for the fermionic $X$ it is odd. This allows  the
following new terms in the MSSM superpotential (model A):
\begin{eqnarray}
W_{new}~=~\lambda_{i} Nu^c_iX+\lambda^\prime_{ij} d^c_id^c_j
\overline{X} + {M_N \over 2} NN+M_X X \overline{X}~.
\end{eqnarray}
Here $i,j$ are family indices with
$\lambda^\prime_{ij}=-\lambda^\prime_{ji}$ and we have suppressed
the color indices. An alternative possibility is to choose $X$ to
have hypercharge $-2/3$ and write a superpotential of the form
(model B)
\begin{eqnarray}
W_{new}~=~\lambda_{j} Nd^c_j X+\lambda^\prime_{kl} u^c_kd^c_l
\overline{X}+{M_N \over 2} NN+M_X X \overline{X}~.
\end{eqnarray}
In model B, additional discrete symmetries are needed to forbid
couplings such as $QL\overline{X}$ which could lead to rapid proton
decay. In model A however, there are no other terms that are gauge
invariant and $R$--parity conserving. In particular, the $X$ field
of model A does not mediate proton decay.  We will illustrate our
mechanism using model A although all our discussions will be valid
for model B as well.

The fermions $N$ and $\tilde{X}$ have masses $M_N$ and $M_X$
respectively.  As for the scalar components of these superfields,
the Lagrangian including soft SUSY breaking terms is given by
\begin{eqnarray}
-{\cal L}_{\rm scalar}\! &=&\! |M_X|^2 (|X|^2 + |\overline{X}|^2) +
m_X^2 |X|^2 + m^2_{\overline{X}}|\overline{X}|^2  \nonumber \\&+&
(B_X M_X X \overline{X} + h.c) + |M_N|^2 |\tilde{N}|^2 +
m_{\tilde{N}}^2 |\tilde{N}|^2  \nonumber \\&+& ({1 \over 2} B_N M_N
\tilde{N}\tilde{N} + h.c.)~
\end{eqnarray}
The $2 \times 2$ mass matrix in the $(X, \overline{X}^*)$ sector can
be diagonalized to yield the two complex mass eigenstates  $X_1$ and
$X_2$ via the transformation
\begin{eqnarray}
X &=& \cos\theta X_1 - \sin\theta e^{-i\phi} X_2; \\
\nonumber \overline{X}^* &=& \sin\theta e^{i\phi} X_1 + \cos\theta
X_2 \end{eqnarray} where
\begin{equation}
 \tan2\theta  = {|2 B_X M_X| \over |m_X^2 -
m_{\overline{X}}^2|};~\phi = {\rm Arg}(B_X M_X) {\rm
sgn}(m_X^2-m_{\overline{X}}^2)~.
\end{equation}
Note that the angle $\theta$ is nearly $45^0$ if the soft masses for
$X$ and $\overline{X}$ are equal.  The two mass eigenvalues are
\begin{equation}
M_{X_{1,2}}^2 \!=\! |M_X|^2 + {m_X^2 + m_{\overline{X}}^2 \over 2}
\pm \sqrt{\left({m_X^2 - m_{\overline{X}}^2 \over 2}\right)^2 + |B_X
M_X|^2 }~
\end{equation}

The two {\it real} mass eigenstates from the $\tilde{N}$ field have
masses
\begin{equation}
M^2_{\tilde{N}_{1,2}} = m_{\tilde{N}}^2 + |M_N|^2 \pm |B_NM_N|~.
\end{equation}
Here $\tilde{N}_1$ is the real part of $\tilde{N}$, while
$\tilde{N}_2$ is the imaginary part. (A field rotation on
$\tilde{N}$ has been made so that the  $B_N M_N$ term is real.) With
these preliminaries we can now discuss baryogenesis and dark matter
in our model.

\section{ Post--Sphaleron Baryogenesis}. The mechanism for generation
of matter-anti-matter asymmetry closely follows the post--sphaleron
baryogenesis scheme of Ref. \cite{bmn}. As the universe cools to a
temperature $T$ which is below the mass of the $X$ particle but
above $M_N$, the $X$ particles annihilate leaving the Universe with
only SM particles and the $N$ (fermion) and $\tilde{N}_{1,2}$
(boson) particles in thermal equilibrium. The decay of $N$ will be
responsible for baryogenesis. We therefore need to know the
temperature at which the interactions of $N$ go out of equilibrium.
We first consider its decay. Being a Majorana fermion, $N$ can decay
into quarks as well as antiquarks: $N \rightarrow u_id_jd_k$, $N
\rightarrow \overline{u}_i \overline{d}_j \overline{d}_k$.  The
decay rate for the former is
\begin{equation}
\Gamma_N \!= \!{C \over 128} {(\lambda^\dagger \lambda) {\rm
Tr}[\lambda'^\dagger \lambda')] \over 192 \pi^3}\sin^22\theta M_N^5
\left({1 \over M_{X_1}^2} - {1 \over M_{X_2}^2}\right)^2
\end{equation}
Here the approximation $M_{X_{1,2}} \gg M_N$ has been made.  $C$ is
a color factor, equal to 6.  The total decay rate of $N$ is twice
that given in Eq. (8) - to account for decays into quarks as well as
into antiquarks. As a reference, we take the contribution from $X_1$
exchange to dominate the decay, and assume that the mixing angle
$\theta \simeq 45^0$. It is then easy to see that for
$\sqrt{(\lambda^\dagger \lambda) {\rm Tr}[\lambda'^\dagger
\lambda')]}\sim 10^{-3}$, $N$ decay goes out of equilibrium below
its mass. Other processes involving $N$ such as $q+N\to
\bar{q}+\bar{q}$ also go out of equilibrium at this temperature.
Further, for $T< M_N$, production of $N$ in $q+\overline{q}$
scattering will be kinematically inhibited. Finally there is a range
of parameters in our model, e.g.,  $M_{\tilde{X}}\sim 3 $ TeV,
$M_N\sim 100 $ GeV, where the rate for $NN\to u^c\bar{u}^c$ process
which occurs via the exchange of the bosonic field in $X$ also goes
out of equilibrium. We have checked that if $N$ decay lifetime is
$\leq10^{-11}$ sec., as it is in our model, even if $NN\to
u^c\bar{u}^c$ is in equilibrium, slightly below $T=M_N$, the decay
rate dominates over this process and does not inhibit baryogenesis.

The decay of $N$, which is CP violating when one--loop corrections
are taken into account, can lead to the baryon asymmetry. Since the
mass of the $N$ fermion is below the electroweak scale, the
sphalerons are already out of equilibrium and cannot erase this
asymmetry. The mechanism is therefore similar to the post--sphaleron
baryogenesis mechanism a la Ref. \cite{bmn}. The only difference
from the detailed model in Ref. \cite{bmn} is that, there, due to
the very high dimension of the decay operator, the out of
equilibrium temperature was above the decaying particle (called $S$
in Ref. \cite{bmn}) mass giving an extra suppression factor of
$T_d/M_S$  in the induced asymmetry (since generation of matter has
to start when the temperature is much below the $S$ particle mass).
In the present case, there is no suppression factor of $T_d/M_S$ in
the induced baryon asymmetry.

In order to calculate the the baryon asymmetry of the universe, we
look for the imaginary part  from the interference between the
tree--level decay diagram and the one--loop correction arising from
$W^\pm$ exchange. These corrections have a GIM--type suppression,
since the $W^\pm$ only couple to the left--handed quark fields while
the tree--level decay of $N$ is to right--handed quarks. Following
Ref. \cite{bmn}, we find the dominant contribution to be
 \begin{eqnarray} {\epsilon_B \over {\rm Br}} \simeq \left\{ -\frac{\alpha_2}{4}\right\}
         {{\rm Im}[(\lambda^* \hat{M_u})^T V \hat{M_d}\lambda'].[\lambda'^*\hat{M_d} V^T \lambda
         \hat{M_u}] \over {M^2_W M_N^2(\lambda^\dagger \lambda) {\rm Tr}(\lambda'^\dagger\lambda')}} \label{asym}
\end{eqnarray}
where ${\rm Br}$ stands for the branching ratio into quarks plus
anitquarks, and $\left(\lambda^*\hat{M_u}\right)^T =
(\lambda_1^*m_u, \lambda_2^* m_c, \lambda_3^*m_t)$, $\hat{M_d} =
diag.\{m_d, m_s, m_b\}$ The interesting point is that as in Ref.
\cite{bmn} the asymmetry is completely determined by the electroweak
corrections. A typical leading term in Eq. (\ref{asym}) is of the
form $(-\alpha_2/4)(m_c m_t m_s m_b)/(m_W^2 m_N^2)$ which yields
$\epsilon_B\simeq 3\times 10^{-8}$ with only mild dependence on the
couplings $\lambda_i, \lambda'_{ij}$. This can easily lead to
desired value for the baryon asymmetry.

\section{ Scalar dark matter}. In a supersymmetric model, we expect
every particle to have a super-partner. We show below that in our
extended MSSM the super-partner of $N$ (denoted by $\tilde{N}_1$)
has all the properties quite naturally for it to play the role of
scalar dark matter. In this context let us recall some of the
requirements on a dark matter candidate: it must be the lightest
stable particle and its annihilation cross section must have the
right value so that its relic density gives us $\Omega_{\rm DM}
\simeq 0.25$. The desired cross section for a generic multi-GeV CDM
particle is of about $10^{-36}$ cm$^2$. In our model the presence of
the TeV scale $X$ particle, in addition to playing an important role
in the generation of baryon asymmetry, also plays a role in giving
the right annihilation cross section for $\tilde{N_1}$ to be the
dark matter.

Let us first discuss why $\tilde{N}_1$ is naturally the lightest
stable boson in our model. To start with, in order to solve the
baryogenesis problem, we choose the $N$ superfield to have its mass
below that of the super-partners of the SM particles. In mSUGRA type
models, generally, one chooses a common scalar mass for all
particles at the SUSY breaking scale (say $M_P$), so that scalar
masses at the weak scale  are determined by the renormalization
group running. There are two kinds of contributions to the running
of the soft SUSY breaking masses -- gauge contributions which
increase masses as we move lower in scale, and Yukawa coupling
contributions which tend to lower the masses as we move lower in
scale. As far as the scalar $\tilde{N}_1$ particle goes since it has
no gauge couplings, its mass naturally goes somewhat lower as we
move from the Planck scale to the weak scale and becomes naturally
the lightest stable SUSY particle. Furthermore since its couplings
$\lambda_{i}$ are in the range of 0.1-0.001, they are not strong
enough to drive $m^2_{\tilde{N}_1}$ negative like the $m^2_{H_u}$.

From Eq. (7), it is clear that of the two states $\tilde{N}_{1,2}$,
the lighter one $\tilde{N}_1$ is the LSP.  The $\tilde{N}_2$ remains
close in mass but above the LSP and can help in co-annihilation of
the dark matter provided $|BM_N| \ll M_N^2 + m^2_{\tilde{N}}$, if
needed.

{\bf Dark matter annihilation}. In the early universe, the LSP
$\tilde{N}_1$ will annihilate into quark-antiquark pair via the
exchange of $\tilde{X}$ fermion.  The annihilation cross section is
given by
\begin{eqnarray}
\! \sigma(\tilde{N}_1 \tilde{N}_1 \rightarrow q \overline{q})v_{\rm
rel}\! = \!{C'(\lambda^\dagger \lambda)^2 \over 8 \pi s}\left({a
\over b} \tanh^{-1}({b\over a})-1\right)
\end{eqnarray}
where
\begin{equation}
       a=2 E^2-M^2_{\tilde{N}_1}+M_X^2;~~~b = 2E|\overrightarrow{p}|~.
\end{equation}
Here $C'=3$ is a color factor, $E$ and $\overrightarrow{p}$ are the
energy and momentum of one of the $\tilde{N}_1$ , $s$ is the total
CM energy. For $M_X \gg E$, the cross section reduces to
\begin{equation}
\sigma v_{\rm rel} \simeq {1 \over 8 \pi} (\lambda^\dagger
\lambda)^2  {|\overrightarrow{p}|^2 \over M_X^4}~.
\end{equation}
For the coupling $ \lambda_3 \sim 1$, $M_{\widetilde{N_1}} = 300$
GeV and $M_X = 500$ GeV, the cross section is of the order of a pb
as would be required to generate the right amount of relic density.

We can now compare this with the dark matter in MSSM, which is
usually a neutralino. In MSSM, some tuning of parameters is needed,
either to have the right amount of Higgsino content in the LSP, or
to have the right--handed stau nearly mass degenerate with the LSP
to facilitate co-anninhilation. In our model, there is no need for
co-annihilation, but if necessary, the mass of $\tilde{N}_2$ is
naturally close to $\tilde{N}_1$ by a symmetry, viz., supersymmetry,
if $|BM_N|$ is small.

{\bf Dark matter detection}. Due to the fact that $\tilde{N}_1$ has
interactions with quarks which are sizeable, it can be detected in
current dark matter search experiments. We present an order of
magnitude estimate of the $\tilde{N}_1$ + nucleon cross section.
Even though the annihilation cross section is of order $10^{-36}$
cm$^2$, the detection cross section on a nucleon
$\sigma_{\tilde{N}_1+p}$ is much smaller due to slow speed of the
dark matter particle which limits the final state phase space for
the elastic scattering. Secondly, detection involves only the first
generation quarks whereas annihilation involves the second
generation as well and thus if the $N$ couplings are hierarchical
like the SM Yukawa couplings, it is easy to understand the smallness
of detection cross sections compared to $\sigma_{ann}$. In our model
the scattering  of $\tilde{N}_1$ (with momentum $p$) off a quark
(with momentum k) occurs via the s-channel exchange of the fermionic
component of $X$ . The amplitude is given by
${\cal M}_{\tilde{N}_1+ q} =
i\frac{\lambda_1^2}{4M_X^2}\overline{u(k')}\gamma^{\mu}u(k)Q_{\mu}$,
where, $Q = k + p$. At the nucleon level , the time component of the
vector current dominates (spin-independent) over the spatial
component (velocity dependent). The nucleon--$\tilde{N}_1$ cross
section is given by
\begin{eqnarray}
\sigma_{\tilde{N}_1+ p}\simeq \frac{|\lambda_1|^4 m^2_p}{64\pi
M^4_{X}}\left( \frac{A  +Z}{A}\right)^2,
\end{eqnarray}
where $A,Z$ are the atomic number and charge of the nucleus and
$\lambda_1$ is the coupling of $N$ to the first generation quarks.
The sum $|\lambda_1|^2+|\lambda_2|^2$ is constrained by LSP
annihilation requirement but individually $|\lambda_1|$ is not. If
we choose $|\lambda_1|\sim 0.1 - 0.01$, then the cross section is
around $10^{-43}\; cm^2 - 10^{-47}\; cm^2$ which is in the range
being currently explored \cite{Xenon-SuperCDMS}.

\section{ Neutron-anti-neutron oscillation}. One of the interesting
predictions of our model is the existence of neutron-anti-neutron
oscillation at an observable rate. The Feynman diagram contributing
to this process is given in Ref. \cite{nnbar}. Since $N$ is a
Majorana fermion, it decays into $udd$ as well as into
$\bar{u}\bar{d} \bar{d}$, which leads to $N-\overline{N}$
oscillations. The strength for this process (taking into account the
anti-symmetry of $\lambda^\prime$ couplings) is given by:
\begin{eqnarray}
G_{\Delta B=2}~\simeq \frac{(\lambda_{1}\lambda^\prime_{12})^2
}{M_NM^4_X}~.
\end{eqnarray}
The $\Delta B=2$ operator in this case has the form
$u^cd^cs^cu^cd^cs^c$.  The coupling $\lambda_1$ appearing in this
process is involves the first generations, the same coupling appears
in the direct detection of dark matter. It is reasonable to expect
$\lambda_1$ to be somewhat smaller in magnitude compared to the
second generation counterpart $\lambda_2$. Secondly, if we choose
the strange quark component in the nucleon to be about $1\%$, then
choosing $\lambda_{1}\lambda^\prime_{12}\simeq 10^{-4}$, we find
that $G_{\Delta B=2}\simeq 10^{-27}$ GeV$^{-5}$ which corresponds to
the present limit on $\tau_{N-\bar{N}}\sim 10^8$ sec.
\cite{milla,nnbar1}. There are proposals to improve this limit by
two orders of magnitude \cite{yuri} by using a vertical shaft for
neutron propagation in an underground facility e.g. DUSEL. It is
interesting that the expectation for the $N-\overline{N}$ transition
is in the range accessible to experiments and this can therefore be
used to test the model.

It is important to point out here that there is no proton decay in
this model due to the fact that both the scalar and the fermionic
parts of the singlet field $N$ are heavier than SM fermions.

$N$ can be identified with the right--handed neutrino, but its
couplings to the light neutrinos are forbidden.  If this model is
embedded into a seesaw picture, we are envisioning a $3\times 2$
seesaw with two heavy right--handed neutrinos and a light one that
is identified with the $N$ field that plays no role in neutrino mass
physics. This can be guaranteed by demanding that $N$ and $X$ fields
are odd under a $Z_2$ symmetry whereas all other fields are even.
The $X\bar{X}$ mass term breaks this symmetry softly and does not
affect the discussion.  Note that proton decay via the exchange of
$N$ is forbidden in this case.

We conclude by noting some interesting aspects of the model.

(i) The $X$ particle in our model can be searched for at the LHC.
Once produced, $X$ will decay into two jets, e.g., a $b$ jet and a
light quark jet.  We point out that there is an interesting
difference in discovery of SUSY at LHC in our model. Consider up
type squarks pair produced at LHC.  The squark will decay into a
quark plus a neutralino. In our model, the neutralino is unstable,
it decays into $u^c d^c d^c \tilde{N}$.  So one SUSY signal will be
six jets plus missing energy.  The scalar up squark can also decay
directly into $\tilde{N} d^c d^c$.  In this case the signature will
be 4 jets plus
 missing energy.\\
(ii) It is also worth noting that the quantum numbers of $N$ are
such that it is a SM singlet with $B-L=1$ and therefore same as that
of the conventional right--handed neutrino. This model can therefore
be used to understand the small neutrino masses via a low scale
seesaw mechanism provided there are at least two $N$'s and the Dirac
masses for neutrinos are suppressed. We do not dwell on this aspect
of the model in this paper since it is not pertinent to our main
results. We however point out that our results are not affected by
the multi--$N$ extension required for understanding neutrino masses.
The RH neutrino which plays the role in generating baryon asymmetry
and dark matter is the lightest of the $N$ fields. This model is
however different in many respects from some other suggestions of
right handed sneutrino dark matter in literature
\cite{Arkani-Hamed,LMN, shri}.\\ $(iii)$ The models presented are
compatible with gauge coupling unification, provided that the $X$
particle is accompanied by other vector--like states which would
make complete $10 + \overline{10}$ representations of $SU(5)$. These
extra particles will have no effect on baryogenesis and dark matter
phenomenology.

$(iv)$ We also note that there is no one loop contribution to
neutron electric dipole moment in our model due to the $\lambda'$ or
$\lambda$ couplings since they involve products of couplings of the
form $\lambda^\dagger\lambda$ and similarly for $\lambda'$. We have
also not found any two loop diagram involving the X or $N$ exchange
that would contribute to neutron edm.


Acknowledgement: The work of KSB is supported by DOE Grant Nos.
DE-FG02-04ER46140 and DE-FG02-04ER41306, RNM is supported by the
National Science Foundation Grant No. Phy-0354401 and  S. Nasri by
DOE Grant No. DE-FG02-97ER41029.

\end{document}